\documentclass[9pt,twocolumn,twoside]{osajnl}

\journal{ao} 

\setboolean{shortarticle}{true}

\ifthenelse{\boolean{shortarticle}}{\colorlet{color2}{color2b}}{\colorlet{color2}{color2}} 

\title{Connections between transport of intensity equation and two-dimensional phase unwrapping}

\author[1,2,*]{Chao Zuo}

\affil[1]{Smart Computational Imaging (SCI) Laboratory, Nanjing University of Science and Technology, Nanjing, Jiangsu Province 210094, China}
\affil[2]{Jiangsu Key Laboratory of Spectral Imaging $\&$ Intelligent Sense, Nanjing University of Science and Technology, Nanjing, Jiangsu Province 210094, China}
\affil[*]{Corresponding author: surpasszuo@163.com}

\dates{Compiled \today}

\begin{abstract}

In a recent publication [Appl. Opt. 55, 2418 (2016)], a method for two-dimensional phase unwrapping based on the transport of intensity equation (TIE) was studied. We wish to show that this approach is associated with the standard least squares phase unwrapping algorithm, but with additional numerical errors.

\end{abstract}
\setboolean{displaycopyright}{true}
\ociscodes{ (100.5088) Phase unwrapping; (100.5070) Phase retrieval; (120.5050) Phase measurement.}

\doi{\url{http://dx.doi.org/10.1364/AO.XX.XXXXXX}}

\begin{document}

\maketitle
\thispagestyle{fancy}

\ifthenelse{\boolean{shortarticle}}{\ifthenelse{\boolean{singlecolumn}}{\abscontentformatted}{\abscontent}}{}

\noindent Transport of intensity equation (TIE), originally derived by Teague \cite{Teague83} from the Helmholtz equation under paraxial approximation, has gained increasing attentions and applications in the fields of adaptive optics, x-ray imaging, electron microscopy, and optical microscopy during the last decades. As an on-interferometric single-beam phase retrieval method, TIE needs only a minimum of two intensity measurements of the complex field at closely spaced planes perpendicular to beam propagation direction. By solving a two-dimensional (2D) second-order elliptic partial differential equation, a \emph{continuous} phase can be directly recovered, and the complexity resulting from the 2D phase unwrapping can be bypassed \cite{Barty98,Zuo13}. Taking advantage of this appealing feature, Pandey \emph{et. al.} \cite{Pandey16} recently reported an interesting extension of TIE by using it as a 2D phase unwrapping approach instead of conventional phase retrieval. By constructing an auxiliary complex field with a wrapped phase profile, an unwrapped phase free from any discontinuities can be recovered by following three basic steps that just like performing a numerical simulation of TIE phase retrieval. Obviously, this approach should work as the quantitative nature of TIE phase retrieval has been verified by numerous simulations among the literature. In fact, this idea is exactly the same as the phase demodulation approach we proposed about 3 years ago \cite{Zuo13OC}.

In this Comment, we will not discuss the reasonability and applicability of TIE for 2D phase unwrapping, which we believe has been thoroughly demonstrated in \cite{Pandey16} and \cite{Zuo13OC}. The point we want to remark here is that this approach is closely related to the standard least squares phase unwrapping algorithm, which has been proposed for more than 20 years \cite{Ghiglia94}. The key idea of the TIE-based approach is to convert the 2D phase unwrapping into a problem of solving a Poisson equation (which is a simplified version of TIE under uniform intensity):

\begin{equation}\label{eq1}
{{\nabla ^2}\phi  = \rho _{TIE}}
\end{equation}
where $\phi$ is the continuous phase to be solved, ${\rho _{TIE}} =  - k{\textstyle{{\partial I} \over {\partial z}}}$, $k$ is the wave-number, $I$ is the intensity which is assumed to be one, ${\textstyle{{\partial I} \over {\partial z}}}$ is the intensity derivative along z, which can be approximated by a finite difference taken between two closely separated images with $\pm \Delta z$ defocus distances
\begin{equation}\label{eq2}
{\frac{{\partial I}}{{\partial z}} = \frac{{{I_{\Delta z}} - {I_{ - \Delta z}}}}{{2\Delta z}}}
\end{equation}
\begin{figure*}[!htbp]
\centering
{\includegraphics[width=\linewidth]{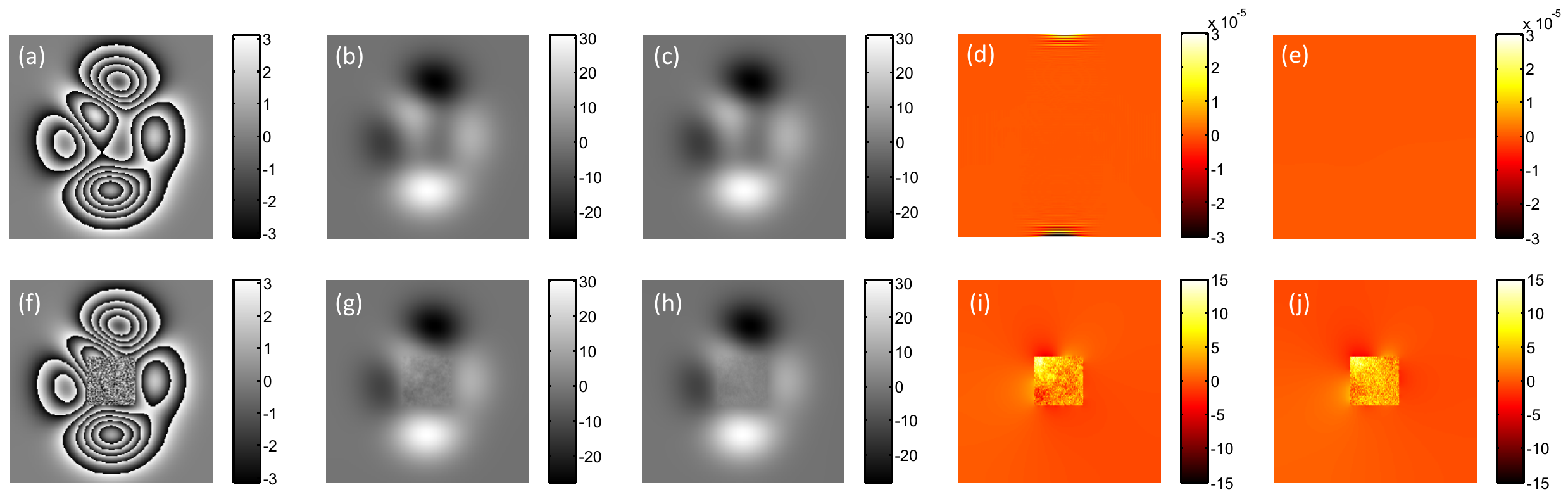}}
\caption{Comparison of the TIE-based approach and the least squares phase unwrapping algorithm. (a) wrapped phase generated from the MATLAB build-in function ‘peaks’; (b) unwrapped phase from (a) with TIE-based approach; (c) unwrapped phase from (a) with least squares approach; (d) error of (b); (e) error of (c); (f) wrapped phase with a small patch containing high density of residues; (g) unwrapped phase from (f) with TIE-based approach; (h) unwrapped phase from (f) with least squares approach; (i) error of (g); (j) error of (h).}
\label{fig1}
\end{figure*}
The two images ${I_{\Delta z}}$ and ${I_{\Delta-z}}$ are artificially generated by numerical propagation based on Fresnel diffraction, or angular spectrum methods. The Poisson equation can be efficiently solved using fast Fourier transform (FFT) \cite{Paganin98} or discrete cosine transform (DCT) \cite{Zuo14}. It was also shown that the FFT-based mirror padding scheme \cite{Volkov02} can be mathematically collapsed into the DCT-based TIE solver under homogeneous (zero) Neumann boundary conditions \cite{Zuo14,Zuo142}.

The least squares phase unwrapping algorithm, however, belongs to the minimum-norm phase unwrapping methods, which formulate the phase unwrapping problem in a minimum-norm sense \cite{Ghiglia94,Ghiglia98}. The minimum-norm methods seek a phase function whose unwrapped phase gradient is as close as possible to the measured wrapped phase gradient. In un-weighted least squares phase unwrapping algorithm, the following $L_2$ norm is used to measure the fitting error
\begin{equation}\label{eq3}
J = {\left\| {\frac{{\partial \phi }}{{\partial x}} - W\left( {\frac{{\partial {\phi _w}}}{{\partial x}}} \right)} \right\|^2} + {\left\| {\frac{{\partial \phi }}{{\partial y}} - W\left( {\frac{{\partial {\phi _w}}}{{\partial y}}} \right)} \right\|^2}
\end{equation}
Where $\nabla  = \left( {\frac{\partial }{{\partial x}},\frac{\partial }{{\partial y}}} \right)$,  ${\phi _w}$ is the wrapped phase, and $W\left(  \cdot  \right)$ is the rewrapping operator. The Euler-lagrange equation of the error function is given by the following Poisson equation:
\begin{equation}\label{eq3}
{\nabla ^2}\phi  = {\rho _{LS}}
\end{equation}
where ${\rho _{LS}} = \nabla  \cdot \left( {W\left( {\frac{{\partial {\phi _w}}}{{\partial x}}} \right),W\left( {\frac{{\partial {\phi _w}}}{{\partial y}}} \right)} \right)$. Thus, the minimization of Eq. (3) boils down to finding the solution of the above Poisson equation. Similarly, this equation can be solved under Neumann boundary conditions using DCT-based Poisson solver \cite{Ghiglia94}, or FFT with a mirror padding scheme \cite{Pritt94}. These Poisson solvers can be viewed as a special inverse filter performed in the Fourier or DCT domain.

Careful comparison of Eq. (1) and Eq. (4) reveals that the only difference between the two approaches lies in their ways to determine the desired phase Laplacian ${\nabla ^2}\phi$. In TIE-Based approach, ${\rho _{TIE}}$ is obtained by the finite difference between two artificially propagated intensities along the axial direction, while in the least squares phase unwrapping algorithm, ${\rho _{LS}}$ is directly calculated using finite differences over the neighbor-hooding pixels. For this reason, there are subtle differences between the filter coefficients in the Fourier or DCT domain (a parabolic inverse Laplacian filter \cite{Zuo13OC,Pandey16} versus a standard discretized Poisson solver \cite{Ghiglia94,Pritt94}).

\begin{figure}[!htbp]
\centering
{\includegraphics[width=\linewidth]{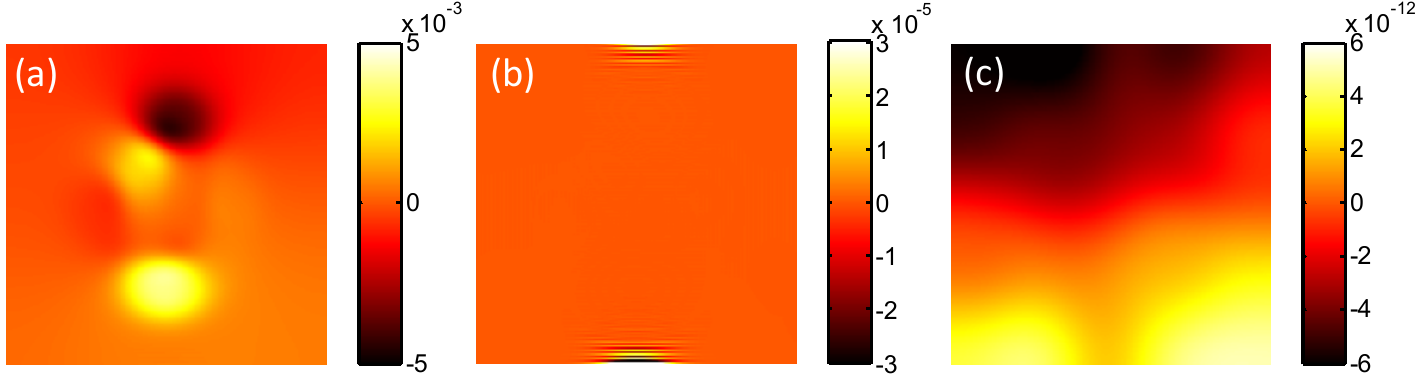}}
\caption{Comparisons of error distributions under ideal conditions. (a) TIE-based approach using angular spectrum method; (b) TIE-based approach using Fresnel diffraction; (c) least squares algorithm. Note that each error distributions are displayed with their respective full dynamic range, and thus, the scale bars are different.}
\label{fig1}
\end{figure}

Since the link between the TIE-based approach and the least squares phase unwrapping algorithm is established, the performance of these two approaches is expected to be similar, which can be verified by our numerical simulations shown in Fig. 1. It can be found that under ideal (noise- and residue-free) conditions, the least squares approach gives a perfect reconstruction (phase error $\sim10^{-12})$, while the numerical error of the TIE-based approach is larger (maximum phase error $\sim10^{-5}$, mainly concentrates on the edge regions), but still negligible. It should also be noted that the two out-of-focus (${I_{\Delta z}}$ and ${I_{\Delta-z}}$ ) images should be better generated by Fresnel diffraction instead of angular spectrum method as the TIE assumes paraxiality. A detailed comparison of the TIE-based approach (using Fresnel diffraction, or angular spectrum method for beam propagation) and the least squares algorithm under ideal conditions are further shown in Fig. 2. As we can see, though both the TIE-based approach and the least squares phase unwrapping algorithm can produce a correct unwrapped output, the latter’s numerical error is much less pronounced.

As demonstrated above, although the TIE-based approach and the least squares phase unwrapping algorithm seek the solutions under different considerations, they finally converge to the same theme. Due to the increased computational cost and larger numerical error compared with the more straightforward least squares method, the introduction of TIE into the field of phase unwrapping seems to be an \emph{apple of Sodom}. But the established connections between the TIE and 2D phase unwrapping may offer new chances to further exchange of ideas between these two seemingly disparate areas of research. For example, in TIE phase retrieval, the presence of intensity zeros (or values close to zero) may lead to strong perturbation and discrepancy in the reconstructed phase. Such tricky problem is often attributed the Teague’s assumption \cite{Schmalz11,Zuo143}, or the non-uniqueness of the TIE solution \cite{Gureyev95}, and still has not been well solved. However, in the field of 2D phase unwrapping, a paralleled issue of how to better accommodate the unreliable data distributed typically around the low-modulation areas [like the case shown in Fig. 1(f)] has already been extensively studied and well addressed. By reformulating the least squares problem [Eq. (3)] into a weighted one, the corrupted regions can be elegantly weighted or rejected so that a more accurate result can be obtained \cite{Ghiglia94,Lu07}. We envisage that this idea should also apply to solving the corresponding problem in TIE phase retrieval, which deserves further investigations.

\section*{Acknowledgments}
C. Zuo thanks the Zijin Star program of Nanjing University of Science and Technology.

\section*{Funding Information}

\textbf{Funding.} National Natural Science Fund of China (NSFC) (11574152, 61505081, 61377003); `Six Talent Peaks' project (2015-DZXX-009); `333 Engineering' research project (BRA2015294); Fundamental Research Funds for the Central Universities (30915011318); Open Research Fund of Jiangsu Key Laboratory of Spectral Imaging \& Intelligent Sense (3092014012200417).

\bibliography{sample}

\ifthenelse{\equal{\journalref}{ol}}{%
\clearpage
\bibliographyfullrefs{sample}
}{}


\ifthenelse{\equal{\journalref}{aop}}{%
\section*{Author Biographies}
\begingroup
\setlength\intextsep{0pt}
\begin{minipage}[t][6.3cm][t]{1.0\textwidth} 
  \begin{wrapfigure}{L}{0.25\textwidth}
    \includegraphics[width=0.25\textwidth]{john_smith.eps}
  \end{wrapfigure}
  \noindent
  {\bfseries John Smith} received his BSc (Mathematics) in 2000 from The University of Maryland. His research interests include lasers and optics.
\end{minipage}
\begin{minipage}{1.0\textwidth}
  \begin{wrapfigure}{L}{0.25\textwidth}
    \includegraphics[width=0.25\textwidth]{alice_smith.eps}
  \end{wrapfigure}
  \noindent
  {\bfseries Alice Smith} also received her BSc (Mathematics) in 2000 from The University of Maryland. Her research interests also include lasers and optics.
\end{minipage}
\endgroup
}{}

\end{document}